\documentclass[lettersize,journal]{IEEEtran}
\usepackage{amsmath,amssymb,amsfonts}
\usepackage{algorithmic}
\usepackage{algorithm}
\usepackage{array}
\usepackage[caption=false,font=normalsize,labelfont=sf,textfont=sf]{subfig}
\usepackage{textcomp}
\usepackage{stfloats}
\usepackage{url}
\usepackage{verbatim}
\usepackage{graphicx}
\usepackage{cite}
\newcommand{\mycaps}{\textsc{SonicCaps}}
\newcommand{\myclap}{\textsc{SonicCLAP}}
\newcolumntype{C}[1]{>{\centering\arraybackslash}p{#1}}

\usepackage{multirow}
\usepackage{xcolor}
\usepackage{makecell}
\usepackage{hyperref}
\usepackage{tikz}
\usepackage{booktabs}
\usepackage{babel}   
\usepackage[T1]{fontenc}
\begin{document}

\title{\mycaps: Large-Scale Diverse and Fine-Grained Captioning for Improved Audio-Retrieval}

\author{\IEEEauthorblockN{
        Zineb Lahrichi\IEEEauthorrefmark{1}\IEEEauthorrefmark{2}
        Marc Ferras\IEEEauthorrefmark{1},
        Gaël Richard\IEEEauthorrefmark{2},
        Geoffroy Peeters\IEEEauthorrefmark{2} \\
    }
    \IEEEauthorblockA{
        \IEEEauthorrefmark{1}Sony CTC, France \\
        \IEEEauthorrefmark{2}LTCI, Telecom Paris, Institut polytechnique de Paris, France
    }
}

\maketitle

\begin{abstract}
Recent advances in audio-language modeling have been driven by large-scale audio captioning datasets. However, existing datasets remain limited by low semantic diversity,
generic descriptions lacking acoustic details, and one-to-one audio-caption mappings that poorly reflect the inherent ambiguity of auditory perception. We introduce \mycaps, a large-scale audio captioning dataset comprising \textasciitilde15M captions paired with \textasciitilde700k audio clips, generated using a multi-modal large language model (Qwen3-Omni) conditioned on both audio and text. To explicitly promote diversity, we generate around 24 captions per audio via structured prompt engineering and few-shot generation, spanning main descriptions, rephrased variants (verbosity, style) and semantic tags. Human evaluation shows that \mycaps \space is rated significantly higher than existing captioning datasets, with fine-grained analyses indicating that our captions are perceived as more descriptive and precise, which strongly correlates with quality judgments. Finally, training CLAP models on \mycaps \space with a multi-caption sampling strategy consistently improves audio retrieval and zero-shot classification, with stronger generalization across public and commercial benchmarks. We release both \mycaps \space and two specialized CLAP models on hugging face: \url{https://huggingface.co/datasets/Zineb/SonicCaps}.

\end{abstract}


\section{Introduction}

Understanding and describing sounds through natural language has become a fundamental challenge in multimodal machine learning. Recent surveys emphasize that progress in this area has been strongly shaped by the availability of large-scale datasets built from heterogeneous audio sources and annotation strategies \cite{wijngaard2025audio,mei2022automated,xu2023beyond}. This growing research area has led to the emergence of a wide range of audio-language tasks, including audio-text retrieval \cite{wu2023large}, audio question answering \cite{lipping2022clotho}, text-to-audio generation \cite{DBLP:conf/icml/LiuCYMLM0P23,evans2025stable}, audio understanding \cite{DBLP:conf/iclr/0001LLKG24, chu2023qwen}, audio editing \cite{wang2023audit,DBLP:conf/nips/UngersbockGLYCW25}, amongst others. As the field evolved, the quality, diversity, and scale of audio-language datasets quickly appeared as fundamental requirements for effective audio-language modeling.

The introduction of FreeSound \cite{font2013freesound} 
and AudioSet \cite{gemmeke2017audio} 
marked a turning point for audio-language research. These resources provided the groundwork for scalable supervision through access to large-scale and diverse acoustic content. Building upon this foundation, subsequent works progressively shifted from raw collection pipelines toward the design of curated benchmarks with improved label taxonomies \cite{Stamatiadis2024}, new online-harvested datasets \cite{chen2020vggsound,wu2023large}, richer linguistic annotations produced either through human annotation \cite{kim2019audiocaps, drossos2020clotho} or automated LLM-based labeling \cite{mei2024wavcaps, bai2025audiosetcaps}, and more balanced category distributions \cite{drossos2020clotho,dixit2026foleybench}.

However, current audio-language datasets exhibit several critical shortcomings. The limited size of sound event libraries and substantial variation across data providers, particularly between freely available and professional audio collections, creates a persistent distributional shift that significantly affects model generalization \cite{chang2024context}. Beyond data distribution issues, existing datasets also lack semantic diversity, which is especially critical for contrastive learning, approaches which rely on large-scale heterogeneous data. Furthermore, a single audio signal can often admit multiple valid textual descriptions and vice versa. Indeed, auditory perception is inherently ambiguous and can be interpreted differently depending on context, focus, and prior knowledge. Such one-to-many mapping is still not exploited in current datasets, which typically provide only one or a few captions per clip. This motivates the investigation of richer captioning strategies that explicitly increase linguistic diversity and better reflect the inherent variability of human perception.

The quality of audio captions presents another major bottleneck in current datasets. In particular, LLM based captioning tends to result in repetitive and overly generic descriptions for acoustically similar events, failing to capture fine-grained variations in sound content. This is especially true for ambiguous sounds such as Foleys, environmental sounds, impact sounds, etc.  LLMs are also likely to propagate biases and output hallucinated descriptions. On the other hand, manual annotation is challenging at scale while introducing inherent human biases and inconsistencies across annotators. More fundamentally, the notion of what constitutes a “good” caption still remains largely undefined. In practice, most existing datasets primarily focus on describing the dominant audio events, often overlooking finer acoustic attributes such as background ambiance, texture, temporal dynamics, or intensity variations, despite their relevance in professional contexts.

\begin{figure*}[t]
    \centering
    \includegraphics[width=1\textwidth]{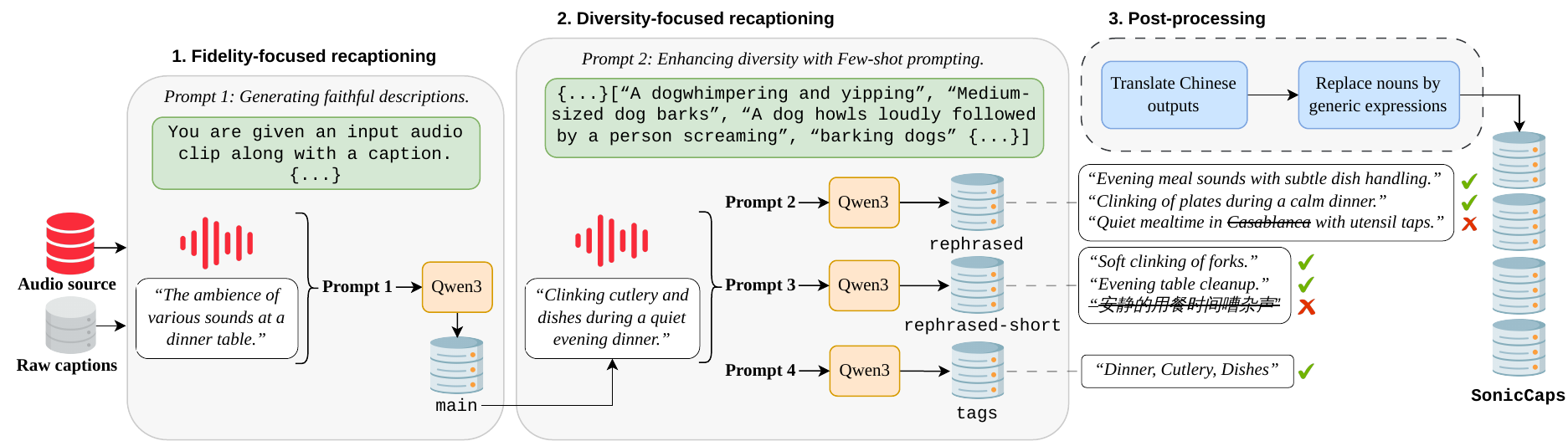} 
    \caption{
    Overview of the proposed caption generation pipeline. High-fidelity captions are first generated from audio, followed by diversity augmentation via prompt engineering and few-shot prompting.}
    \label{fig:main}
\end{figure*}

In this paper, we attempt to bridge the gap between publicly available and commercial captioning data, while encouraging more diverse captioning across heterogeneous audio sources. To this end, we make several contributions:
\begin{itemize}
    \item We release \mycaps, a large-scale audio captioning dataset comprising \textasciitilde15M captions paired with \textasciitilde700k audio clips\footnote{Only captions and their corresponding audio identifiers are released.}. Captions are generated using a multi-modal large language model (\texttt{Qwen3-Omni}) \cite{chu2023qwen} through a three-stage pipeline consisting of \textit{fidelity-focused} recaptioning, \textit{diversity-focused} recaptioning, and post-processing, as illustrated in ~\autoref{fig:main}.
    To reflect the one-to-many nature of auditory perception, we introduce a diverse captioning strategy that generates around 24 captions per audio clip through prompt engineering and few-shot generation, spanning 
    multiple styles and levels of granularity.

    \item We further show that sampling diverse captions per audio from \mycaps \space during training yields consistent improvements on audio retrieval and zero-shot classification tasks over existing baselines. These results highlight the benefits of large-scale caption augmentation and show that carefully balancing the composition of captioning data can significantly enhance performance. These gains hold across both public and proprietary benchmarks, demonstrating strong generalization. We release our best-performing model for the audio retrieval task, \myclap$_{\text{AR}}$.

    \item We introduce a new human evaluation framework for pairwise comparison of captions along several qualitative dimensions. We conduct a subjective human study based on this framework, collecting structured qualitative feedback for each rating.
    We show that our \textit{fidelity-focused} captions consistently align better with human preferences and are perceived as more descriptive, capturing richer acoustic details. Training a CLAP model on these captions improves alignment between CLAP scores and human MOS scores. We release this model as \myclap$_{\text{MOS}}$.

\end{itemize}
We hope that \mycaps \space and \myclap \space  will contribute to advancing research in audio-language multimodal learning, and serve as a foundation for the development of higher-quality models in this domain.

\section{Related Works}

\subsection{Audio Language Datasets}

To address the limited availability of paired audio-text data, a major line of research has focused on curating datasets of human-written captions to support audio-language tasks. Early benchmarks such as AudioCaps (46k {a/c} \footnote{a/c: audio/caption} pairs) \cite{kim2019audiocaps} and Clotho (25k a/c pairs) \cite{drossos2020clotho} introduced manually annotated captions collected from AudioSet and FreeSound audio sources, respectively. However, human annotation remains difficult to scale due to its high cost and variability across annotators, particularly for ambiguous or specialized sound events. As a result, these datasets remain relatively limited in scale and diversity, restricting the generalization capabilities of audio-language models.

In parallel, several large-scale datasets based on automated collection pipelines have emerged. For example, VGGSound (200k a/c pairs) \cite{chen2020vggsound} leveraged YouTube videos to build a large-scale audio-visual benchmark with visually grounded event labels, although limited to classification categories rather than natural language descriptions. More recently, large language models have increasingly been adopted for annotation and dataset augmentation. 
For example ClothoV2\_GPT (100k a/c pairs) \cite{Primus2023} extends the Clotho dataset with GPT-3.5-generated descriptions. 
At a larger scale, LAION-630k (630k a/c pairs) \cite{wu2023large} aggregates data from eight online sources and uses a pretrained T5 model to synthesize captions for samples with only tags or label annotations. 
However, the remaining raw captions are highly noisy, and only a portion of the dataset is publicly available due to licensing restrictions.

WavCaps (400k a/c pairs) \cite{mei2024wavcaps} mitigates noisy annotations via Chat-GPT-assisted captioning and filtering strategies over heterogeneous audio sources, demonstrating that enhancing caption quality induces significant performance gains in multiple downstream audio language learning tasks. 
However, because caption generation remains purely text-conditioned, WavCaps relies on aggressive filtering to mitigate noisy metadata, discarding a substantial fraction of the available data. Moreover, without direct conditioning on the audio, there is no guarantee that captions generated from existing descriptions accurately describe the corresponding audio sample. 
Despite these limitations, WavCaps remains widely adopted in the literature, although, as shown later in this paper, its caption quality remains limited.

Other works leverage audio-language models to generate audio-aware captions for audio-language systems. 
AF-AudioSet (600k a/c pairs) \cite{kong2024improving} uses Audio Flamingo \cite{DBLP:conf/icml/KongGBPVC24} to recaption AudioSet and employs LAION CLAP \cite{wu2023large} scores to filter low-quality audio-text pairs. By varying the CLAP threshold, the authors study the trade-off between dataset size and caption quality. They show that smaller but higher-quality synthetic datasets can lead to comparable or improved text-to-audio generation performance. Auto-ACD (1.5M a/c pairs) \cite{sun2024auto} combines audio-language and audio-visual models to extract contextual information such as detected objects, captions, and labels, which are then integrated by an LLM into a final audio description. Sound-VECaps (1.6M a/c pairs) \cite{yuan2025sound} uses a similar framework but generates captions enriched with temporal, spatial, and environmental context. In contrast, Auto-ACD intentionally restricts visual input to the middle video frame and ignores purely visual content. This additional visual context enables the generation of more detailed descriptions and yields improved audio retrieval performance, but relies heavily on video metadata. More recently, AudioSetCaps (6M a/c pairs) \cite{bai2025audiosetcaps} combines audio-language models for content extraction, LLMs for caption generation, and CLAP-based refinement to improve caption quality. 

While these methods produce more informative captions and improve downstream performance, they primarily focus on caption quality or audio data scaling rather than diversity. In addition, they remain confined to relatively homogeneous distributions, largely centered around AudioSet and VGGSound. 

\subsection{Human Perception based models}

There is a growing interest in developing models that better align with human perception. This has led to efforts to improve the correlation between CLAP-based scores and subjective human judgments, as well as to incorporate human evaluation protocols and comparative analyses of captioning datasets \cite{mei2024wavcaps}. Additionally, \cite{hegde2025aligning} presents a reinforcement learning framework for audio captioning that aligns the model output with human preferences using a custom reward model trained on preference data. In parallel, several works have moved toward more professional and application-driven representations of sound.
For instance, \cite{sridhar2026audiocards} introduces structured metadata tailored to sound designers. Similarly, benchmark datasets grounded in professional sound taxonomies, such as FoleyBench \cite{dixit2026foleybench}, provide more fine-grained and industry-relevant evaluation frameworks.
In this paper, we propose a CLAP model that aligns better with human 
preferences, by training it on our \textit{fidelity-focused captions}, which are shown to be more faithful and perceptually detailed.

\section{\mycaps}

\subsection{Data sources}
\label{section:data-sources}

In this section, we describe the different audio-language sources used to construct \mycaps, including automatic and manually annotated captions. There might be audio overlap across sources, but no overlap over captions.

\textbf{FreeSound} \cite{font2013freesound} is a collaborative, community-driven audio repository introduced for artists and audio researchers. It operates as an open and continuously growing database where any user can upload, download, and re-purpose a wide variety of sounds. It contains more than \textasciitilde 700k audio recordings under Creative Commons licenses (\textasciitilde515k currently harvested for audio-language research), spanning diverse categories, including music samples, environmental sounds and synthesized effects. Each clip is annotated by contributors, with descriptions and tags which can be highly noisy: many are only loosely related to the audio, lack grammatical structure, or include overly specific and non-generalizable details. Additionally, redundancy across user uploads further degrades annotation quality, as the same caption is often associated with different audio clips. 

\textbf{AudioCaps} \cite{kim2019audiocaps} is built upon AudioSet \cite{gemmeke2017audio}, a large-scale corpus for audio event analysis. AudioCaps constitutes the largest benchmark with human-written descriptions for audio captioning. The dataset contains roughly 50k audio excerpts. Audio sources are sampled from an ontology of 527 audio event categories derived from AudioSet, while the number of instances per category is capped at 2k in order to reduce class imbalance. Each training example is paired with a single textual annotation, whereas samples in the validation and test partitions are associated with five independent human-written references. 

\textbf{BBC Sound Effects} The BBC Sound Effects Archive~\footnote{https://sound-effects.bbcrewind.co.uk/} is a collection of over 33k audio recordings gathered over the past century for personal, educational, and research use. Originally developed to support BBC productions, the archive spans a wide range of natural, urban, mechanical, and atmospheric sounds from around the world, including historically significant recordings dating back to 1889. The captions are generally informative, but often exhibit grammatical inconsistencies or incomplete formulations, and frequently include non-generic details such as city names, landmarks, or dates. 

\textbf{WavCaps} is the first large-scale audio captioning dataset, using ChatGPT to scale to over 400k audio-caption pairs sourced from FreeSound, BBC Sound Effects, AudioSet Strongly-Labeled \cite{hershey2021benefit}, and SoundBible \footnote{https://soundbible.com/}.
Since their approach uses unimodal text-to-text generation conditioned on raw captions, the authors apply pre-filtering to remove irrelevant and over-represented descriptions, reducing noisy conditioning signals. However, this filtering discards approximately half of the FreeSound data, potentially limiting the dataset's utility for downstream audio representation learning. Moreover, low-quality raw captions lead to erroneous conditioning, propagating errors into the generated captions. While the filtered captions show improved quality over raw captions, they often lack descriptive detail and exhibit limited diversity across datasets, with frequent grammatical patterns and overly generic descriptions (e.g., “A sound is played”), leaving substantial room for improvement.


\begin{table*}[t]
\centering
\caption{Prompts used for Qwen-based recaptioning.}
\label{tab:qwen_prompts}

\renewcommand{\arraystretch}{1.4}
\setlength{\tabcolsep}{4pt}

\begin{tabular}{c|p{0.82\linewidth}}
\toprule
\textbf{Caption Type} & \textbf{Prompt} \\
\midrule

\texttt{main} &
You are given an input audio clip along with a caption. Use them to generate an event-based caption of the audio content. The new caption must consist of at most one sentence and should not start with an article (A, An, The), but rather with a verb in the -ing form, a noun, or an adjective. Write between 2 and 20 words. Describe what happens in the audio. Do not use a list of tags or keywords, and do not say ``the audio'' or ``the clip''. Do not use ``is heard'' or ``we hear''. Remove author names, locations, city names, country names, time references, and device names. Add punctuation. Describe ambient and background sounds if present. Do not transcribe speech. Indicate the presence of speech, speaker gender, language, and speech properties only if speech is the primary focus. Do not include extra text or metadata. Avoid speculation and describe only clearly audible content.\\
\midrule

\texttt{rephrased} &
You are given an input audio clip along with its caption. Use them to generate ten alternative captions that differ from the original one. Shorter descriptions are allowed. Use ``::'' to separate the generated captions. Use the following examples as diverse formulations of the same concept ``dog barking'' : ``A dog whimpering and yipping'' :: ``Medium-sized dog barks'' :: ``A dog howls loudly followed by a person screaming'' :: ``Dog barks, bees buzz, and rail transport moves'' :: ``barking dogs'' :: ``A door creaks while a dog barks in the distance''. \\
\midrule

\texttt{rephrased-short} & You are given an input audio clip along with its caption . Use them to generate ten different captions that are different from the original. Provide very short descriptions (a few words). Use ``::'' to separate the ten output descriptions.
 \\
 
\midrule
\texttt{tags}  & You are given an input audio clip along with its caption. Use them to generate a list of at most three tags. Tags must be nouns rather than verbs. Use commas to separate the tags.  \\
\bottomrule

\end{tabular}
\end{table*}

\subsection{Multi-modal Audio Captioning}

In this section, we describe the proposed caption generation pipeline, illustrated in \autoref{fig:main}. The pipeline comprises three stages leading to four complementary caption types (see \autoref{tab:qwen_prompts} which summarizes the prompting strategies employed).
The first category, referred to as \textit{fidelity-focused} or \texttt{main} captions, consists of factual, descriptive, and perceptually grounded descriptions that aim to remain as faithful as possible to the audio content (see section \ref{subsubsec-fidelity}.1). The second stage, leading to three additional captions types, referred to as \textit{diversity-focused}, produces multiple variants spanning different styles, verbosity levels, and granularities through few-shot prompting  (see section \ref{subsubsec-fidelity}.2). 
Finally, a \emph{post-processing} stage removes artifacts and consolidates the resulting captions to construct \mycaps.
The prompts presented in \autoref{tab:qwen_prompts} were empirically derived through an iterative prompt engineering process involving trial-and-error refinement. Specifically, we refined the prompting strategies based on subjective evaluations conducted on small validation batches of approximately 50 examples. This procedure enabled progressive adjustments to the prompt formulations, ultimately reaching configurations that consistently    produced high-quality and stable outputs.

\subsubsection{Fidelity-focused recaptioning}
\label{subsubsec-fidelity}
To enrich the linguistic diversity and descriptive quality of the captions, we performed automatic recaptioning using the multi-modal large language model \texttt{Qwen/Qwen3-Omni-30B-A3B-Instruct}. The model was conditioned on audio inputs resampled to 16\,kHz and truncated to a maximum duration of 10 seconds for inference. Caption generation was performed using stochastic decoding with a temperature of 0.6, top-\(p\) sampling with \(p=0.95\), top-\(k=20\), and a maximum generation length of 30 tokens per caption. In practice, we observed that purely audio-conditioned captioning occasionally produced hallucinated or weakly grounded descriptions, particularly for acoustically ambiguous sound events. 
To mitigate this issue, we explored prompt engineering strategies and multi-modal contextualization to encourage captions that remain faithful to the underlying audio content while retaining a high level of descriptive detail. 

In particular, we iteratively refined the prompt as a set of explicit \textit{dos} and \textit{don'ts} to mitigate recurrent generation artifacts and improve both faithfulness and linguistic diversity. First, we discourage generic caption templates commonly produced by Qwen (e.g., captions beginning with \textit{``the audio''}, \textit{``the clip''}, \textit{``we hear''}, or articles such as \textit{``A, An, The''}). Instead, we explicitly instruct Qwen to start captions with a verb in the \textit{-ing} form, a noun, or an adjective. Second, we promote factual grounding by framing the task as \textit{``event-based''} caption generation and instructing the model to describe only clearly audible events while avoiding speculation. Additionally, we explicitly instruct the model to include relevant \textit{``ambient and background sounds when present''}, as these contextual details are frequently missing from existing captioning datasets. Finally, we encourage informative yet concise descriptions to minimize the irrelevant verbosity and noise often introduced by LLMs. Specifically, we constrain caption length both in the prompt (20 words) and during decoding (30 tokens). We further exclude undesirable content such as speech transcriptions, recording metadata, and other information not directly supported by the audio signal. 

We emphasize that these instructions only act as soft constraints rather than strict guarantees, but empirically we found that they effectively reduce undesired generation patterns and improve caption faithfulness.

\subsubsection{Diversity-focused recaptioning}

This stage comprises 3 categories of captions: \texttt{rephrased}, \texttt{rephrased-short} and \texttt{tags}. 
First, in the \texttt{rephrased} setting, we aim to generate approximately ten rephrased captions per audio with high linguistic diversity while maintaining a length distribution comparable to the \texttt{main} captions. 
To improve generation efficiency, the captions are produced jointly within a single sampling forward pass using a fixed seed. 
We further employ \emph{few-shot prompting}, a technique where the model is conditioned on a small number of output references to guide its behavior at inference time. Specifically, we provide stylistically diverse captions describing the same semantic concept, encouraging variation in linguistic expression. These examples are adapted from the survey \cite{wijngaard2025audio}, which reports the first occurrence of the word ``dog'' and by extension, the underlying concept of ``dog barking'' for each captioning dataset examined. The resulting examples illustrate how the same concept can be expressed through different annotation styles.

Generation is conditioned on both the \texttt{main} caption and the corresponding audio input.

The \texttt{rephrased-short} setting is designed to produce concise captions that resemble short user queries or prompts. Approximately ten short captions are generated per sample, conditioned on the \texttt{main} caption and the audio signal.
Finally, the \texttt{tags} setting generates three-word descriptions intended to approximate high-level semantic categories. The generated outputs are encouraged to contain nouns rather than verbs in order to emphasize semantic entities over actions. As in the previous settings, generation is conditioned on both the \texttt{main} caption and the audio input.

\begin{table}[t]
\centering
\caption{Example from FreeSound comparing raw annotations, WavCaps captions, and the four caption types in \mycaps. Multiple captions are separated by ``::''.}
\label{tab:qwen_example}
\renewcommand{\arraystretch}{1.4}
\setlength{\tabcolsep}{4pt}

\begin{tabular}{c|p{5.5cm}}
\Xhline{1pt}
\textbf{Caption Type} & Caption\\
\hline
Raw caption & Turning the pages of the book Ruth in the bible (dutch translation) the church has given my father in 1942. \\
\hline
WavCaps & Pages of a book are being turned.\\
\hline
\texttt{main} & Turning pages of a book with crisp paper sounds in a quiet environment.\\
\hline
\texttt{rephrased} & Crisp turning of paper pages, likely from a book or magazine :: Quiet room with paper shuffling sounds :: Close-up of someone flipping through a thick book :: Page turning sounds in a calm, still environment :: Paper being turned one by one, soft and distinct :: Reading in silence with audible page flips :: A book being read aloud, with crisp page turns :: Someone skimming through a document or notebook :: Page turning with slight crinkling sound :: Paper rustling, possibly browsing through a photo album \\
\hline
\texttt{rephrased-short} & Crisp paper rustling :: Page turning sound :: Book flipping :: Quiet reading noise :: Paper shuffling :: Reading in silence :: Soft page turns :: Book handling sounds :: Still environment :: Paper manipulation\\
 \hline
\texttt{tags} & book, page, paper \\
\Xhline{1pt}
\end{tabular}
\end{table}

\begin{table*}[t]
\centering
\caption{Comparison of dataset statistics. “Raw” denotes the original captions associated with each audio source dataset. The statistics for \mycaps \space are computed on the \textit{fidelity-focused} captions, i.e the \texttt{main} subset.}
\label{tab:dataset_stats}
\renewcommand{\arraystretch}{1.3}

\setlength{\tabcolsep}{2pt}

\begin{tabular}{l |c|c|C{1.2cm}| C{1.4cm} |C{8.5cm} |C{1.2cm} }
\Xhline{1pt}

\textbf{Audio source} &
\textbf{Captions} &
\textbf{\#Audios} &
\textbf{\%Unique Cap.} &
\textbf{Avg \#Words} &
\textbf{Most Frequent Caption (MFC)} &
\textbf{\# Occ. (MFC)}\\

\hline

\multirow{3}{*}{FreeSound}

& Raw
& 515K 
& 46.7
& 10.4 $\pm$ 8.5
& Single note sampled from an analog synthesizer by Modular Samples. 
& 36,773 \\

& WavCaps
& 250K 
& 80.4
& 6.7 $\pm$ 2.6
& Music is being played. 
& 442 \\

& \mycaps
& 515K 
& 92.4
& 11.1 $\pm$ 2.9
& Playing a single sustained note on an analog synthesizer. 
& 317 \\

\hline

\multirow{2}{*}{BBC Sound Effects}

& WavCaps
& 31K 
& 84.5
& 8.5 $\pm$ 4.1
& A clock is striking.
& 462 \\

& \mycaps
& 31K 
& 86.0
& 9.6 $\pm$ 5.0
& A clock is striking.
& 166 \\

\hline

\multirow{2}{*}{AudioSet SL subset}

& WavCaps
& 108K
& 87.6
& 9.7 $\pm$ 3.6
& Music is playing. 
& 2,045 \\

& \mycaps
& 108K
& 91.7
& 11.5 $\pm$ 3.0
& Music plays while a man speaks.   
& 83 \\

\hline

\multirow{2}{*}{AudioSet subset}

& AudioCaps
& 49K
& 91.7
& 8.7 $\pm$ 4.2
& A man speaking. 
& 80 \\

& \mycaps
& 49K
& 97.6
& 11.2 $\pm$ 3.2
& A man speaks while typing on a keyboard. 
& 35 \\

\Xhline{1pt}
\end{tabular}
\end{table*}

\begin{table}[t]
\centering
\renewcommand{\arraystretch}{1.2}

\caption{Statistics of \mycaps}
\label{tab:stats-soniccaps}
\setlength{\tabcolsep}{3pt}
\begin{tabular}{l C{1.5cm} C{1.3cm} C{1.4cm} C{1.2cm}}
\Xhline{0.5pt}
\toprule
Method &  \textbf{\% Unique Caps.} &  \textbf{Avg \#Words} & \textbf{Avg \# Cap.} / \textbf{Audio} & \textbf{Vocab.} \\
\midrule
\texttt{main}  &  92.4 & 11.2 $\pm$ 2.9 & 1 $\pm$ 0.0   & 26K\\
\texttt{rephrased}  & 86.9 & 9.1 $\pm$ 2.7 & 10 $\pm$ 0.2  & 41K \\
\texttt{rephrased-short}& 52.2 & 4.5 $\pm$ 1.7 & 9.9 $\pm$ 0.4  &47K \\
\texttt{tags} & 50.8 & 3.3 $\pm$ 0.7  & 2.9 $\pm$ 0.2  &18K\\
\hline
\mycaps & 61.3 & 5.6 $\pm$ 3.4 & 23.9 $\pm$ 0.5 & 66K \\
\bottomrule
\Xhline{0.5pt}
\end{tabular}
\end{table}

\subsubsection{Post-processing}
Occasionally, some generated captions contained Chinese text, which we translated into English using \texttt{facebook/nllb-200-1.3B}.
In addition, we detected captions containing proper nouns, such as city or country names, and prompted Qwen to regenerate them while explicitly prohibiting the identified entities. Initially, we explored additional filtering strategies to further improve caption quality, including selecting between WavCaps and Qwen-generated captions, as well as sampling multiple generations with different random seeds and retaining the best candidates. However, these approaches were ultimately discarded due to their substantial computational overhead relative to the expected gains. Furthermore, reliably evaluating caption quality remains challenging, as existing automatic metrics are limited. In particular, as discussed in Section \ref{section:results_subjective_test}, the LAION-CLAP score, despite being widely used to assess audio-text alignment, does not reliably reflect perceived caption quality.

\subsection{Dataset overview}

\mycaps \space contains \textasciitilde700k audio clips and \textasciitilde15M audio-caption pairs derived from the four audio sources previously introduced in \autoref{section:data-sources}. Each source is further partitioned into four captioning subsets, corresponding to the four caption types described in \autoref{tab:qwen_prompts}. A representative example from FreeSound is shown in~\autoref{tab:qwen_example}, comparing the raw caption with the corresponding WavCaps and \mycaps \space captions. This example shows how the \texttt{main} caption from \mycaps \space provides richer acoustic details than the baselines. The \texttt{rephrased} caption further increases linguistic diversity through variations in syntax, voice (active/passive), and vocabulary, while the \texttt{rephrased-short} caption provides concise captions resembling user queries. Finally, the \texttt{tags} encode the semantic content of the audio using three keywords.

\autoref{tab:dataset_stats} summarizes dataset statistics for \mycaps \space and existing datasets. We first quantify caption redundancy, reporting both the proportion of unique captions and the most frequent caption within each corpus. This issue is particularly pronounced in FreeSound, where many users upload large collections of sounds sharing identical descriptions. WavCaps reduces this redundancy by filtering high-frequency captions assumed to be weakly correlated with the audio content. However, this filtering step sacrifices almost half the available FreeSound data (\textasciitilde250k against \textasciitilde515k). Additionally, because caption generation remains primarily text-conditioned in WavCaps, similar metadata frequently yields nearly identical descriptions, limiting linguistic diversity. Joint conditioning on audio and text alleviates this effect, resulting in a substantially higher proportion of unique captions across all audio sources in \mycaps. Residual redundancy nevertheless remains for acoustically similar samples sharing the same description, such as collections of hi-hat sounds differing only subtly in texture. This observation suggests that the temporal and spectral resolution of \texttt{Qwen3-Omni} is still insufficient to reliably capture such fine-grained acoustic variations.

Finally, we also report the average caption length in number of words. On average, \mycaps \space produces slightly longer captions than AudioCaps and WavCaps, suggesting richer and more detailed descriptions. \autoref{tab:stats-soniccaps} also reports these statistics for each subset of \mycaps. The \texttt{main} subset contains the longest and most detailed captions.
Caption length then decreases progressively across subsets, from \texttt{rephrased}
to \texttt{rephrased-short},
 and finally to \texttt{tags}.

The table also reports vocabulary size, showing that \texttt{rephrased} and \texttt{rephrased-short} substantially expand the vocabulary compared to \texttt{main} (41K and 47K unique words vs. 26K, respectively). This confirms the role of our \textit{diversity-focused} rephrasing strategy in increasing linguistic diversity.
The LLM consistently generated the expected number of captions per audio for each prompt, with only minor deviations from the target, yielding approximately 24 captions per audio overall. 
Finally, \texttt{main} exhibits the highest caption-level diversity (92.4\% unique), closely followed by \texttt{rephrased} (86.9\%), confirming that our few-shot prompting strategy promotes diverse outputs even when generating 10 captions per input. Diversity drops to 52.2\% for \texttt{rephrased-short} and \texttt{tags}, consistent with their shorter, more generic, query-like nature.

\section{Experiments}

\subsection{Audio Language Retrieval}
\label{subsection:AR-retrieval}

We evaluate the quality of our captioning dataset on an audio-language retrieval task. 
The task consists of retrieving audio clips from text queries and vice versa by learning a shared embedding space via contrastive learning.
We train CLAP-style models following the architecture and training strategy from \cite{DBLP:journals/corr/abs-2604-01929}, using a RoBERTa-Large text encoder \cite{liu2019roberta} and a PaSST audio encoder \cite{DBLP:conf/interspeech/KoutiniSEW22}. The encoder architecture is kept fixed across all experiments while the weights are independently optimized for each model, as investigating alternative CLAP architectures is beyond the scope of this work. A linear projection head maps the pooled audio and text embeddings into a shared 1024-dimensional representation space. Audio signals are resampled to 32 kHz to match PaSST's native sampling rate, and training uses random 10-second excerpts from full recordings. We optimize a symmetric contrastive objective \cite{radford2021learning}, which encourages aligned audio–text pairs to have higher similarity than all other pairs within the batch.
We set the temperature hyperparameter $\tau$ to 0.2. Training is performed over 4 GPUs, with a batch size of 112, yielding an effective batch size of $N=448$. 

We first train a model on a baseline audio-language dataset, denoted as (*), composed of heterogeneous caption sources: AudioCaps (AC), the original FreeSound captions (FS$_{\text{raw}}$), and the BBC Sound Effects and AudioSet Strongly Labeled subsets annotated with WavCaps captions, (BBC+ASL)$_{\text{WC}}$. We exclude SoundBible due to licensing restrictions, though it was originally part of WavCaps. Additionally, since WavCaps does not annotate the full FreeSound corpus, we retain FS$_{\text{raw}}$ captions to maximize audio coverage while keeping the underlying audio data fixed across experimental conditions. This choice is consistent with prior work \cite{bai2025audiosetcaps} showing that CLAP performance is strongly driven by the scale of audio training data. For public release of the CLAP models, we further restrict FreeSound samples to compatible licenses, excluding \textit{cc-by-nc}, \textit{cc-by-nc-sa}, \textit{cc-by-nd}, and \textit{cc-by-nc-nd} content, resulting in 371k audio samples compared to the 250k available in WavCaps.

We then train several CLAP variants using the same underlying audio datasets, while varying the composition and diversity of caption sources through different sampling strategies. For each model, captions are sampled from the baseline dataset and the four \mycaps\ subsets introduced in \autoref{tab:qwen_prompts}, according to a predefined probability distribution. For datasets containing multiple captions per audio sample, we first randomly shuffle the caption set and then uniformly sample a caption from the resulting pool. In addition, we randomly remove punctuation marks with a probability of 0.2 during training.

\begin{table}[t]
\centering

\caption{Taxonomy of observations: Recurring categories of errors observed in audio captions during dataset inspection.}
\label{tab:taxonomy_obs}
\renewcommand{\arraystretch}{1.4} 

\begin{tabular}{p{0.25\linewidth} p{0.65\linewidth}}
\toprule
\textbf{Qualitative observations} & \textbf{Description} \\
\midrule

Lack of descriptive details & Insufficient use of meaningful qualitative information such as adjectives, adverbs, texture, rhythm, intensity, or background context. \\
\hline

Element(s) heard but not described & One or several sound events is heard by the participant but not mentioned in the caption.   \\
\hline

Element(s) described but not heard / not plausible &  One or several sound events are described in the caption but do not match the audio content. Given the inherent ambiguity of auditory perception, multiple valid interpretations may exist; consequently, plausible alternative descriptions (e.g., “hot coffee” vs. “lemon juice” for a pouring sound) are not considered erroneous.\\
\hline

Minor imprecisions & The caption exhibits temporal, numerical or categorical inaccuracies such as incorrect event counts or duration estimates. \\
\hline
Globally unrelated & The caption is semantically inconsistent with, or unrelated to, the audio content. \\
\bottomrule
\end{tabular}
\end{table}

We compare our models against \texttt{LAION-CLAP HTSAT-base} \cite{wu2023large}, which employs RoBERTa and HTSAT encoders and has a comparable number of parameters to our architecture. We further train a reference model on the AudioCaps and WavCaps (AC+WC) datasets using the same audio and text encoders and training strategy as our proposed models, to ensure a controlled comparison of our datasets under identical optimization conditions.

We evaluate the performance of our CLAP models against our baselines on text-to-audio (T2A) and audio-to-text (A2T) retrieval tasks. Models are evaluated using recall $R@k$ with $k \in \{5,10\}$, measuring the fraction of queries for which the ground-truth item appears within the top-$k$ retrieved results. Evaluation is conducted on the AudioCaps validation set, as well as on a mixture of three internal sound effects datasets, from which we extract 500 audio-caption pairs each. These internal datasets are selected to exclude tag-based annotations and retain only natural language sentence-level descriptions. Since the AudioCaps validation set provides five reference captions per audio, we report both $R@k$-any and $R@k$-all. $R@k$-any considers a retrieval successful if at least one reference caption is retrieved, whereas $R@k$-all requires that all five captions are retrieved within the top-$k$ results.

Finally, we investigate the relationship between retrieval performance and caption diversity. 
Since our models are trained by sampling captions from multiple annotation sources, we characterize the effective diversity of the training captions through the \emph{caption sampling perplexity}. 
Let $S$ denote the number of caption sources. Source $i$ is sampled with probability $p_i$, where $\sum_{i=1}^{S} p_i = 1$, and provides $N_i$ candidate captions for each audio sample. Assuming uniform sampling among captions within a source, the probability of selecting an individual caption is $p_i/N_i$. We therefore define the caption sampling perplexity as

$$
\mathrm{PPL}_{\mathrm{captions}}
=
\exp\!\left(
-\sum_{i=1}^{S}
p_i
\log\!\left(\frac{p_i}{N_i}\right)
\right).
$$

This quantity corresponds to the effective number of captions that may be observed for a given audio sample during training. It equals one when training always relies on a single caption and increases as the caption sampling distribution becomes more diverse. We emphasize that this metric captures only the diversity induced by the sampling strategy and caption multiplicity. It does not account for the semantic diversity of the captions themselves, as this quantity cannot be estimated consistently for external baselines for which training data are unavailable.

\begin{table*}[t]
\centering
\caption{Text-to-Audio (T2A) and Audio-to-Text (A2T) performance and ablation study on training caption datasets. Sampling probabilities indicate the proportion of captions drawn from either our baseline audio-language dataset (*)
, or from our proposed \mycaps \space captions (\texttt{main} (M), \texttt{rephrased} (R),  \texttt{rephrased-short} (RS) and \texttt{tags} (Tag)).}
\label{tab:AR_results}
\setlength{\tabcolsep}{3.5pt}

\begin{tabular}{l c | ccccc | cc cc cc | cc cc}
\toprule
\textbf{Model} & \textbf{Training Datasets}
& \multicolumn{5}{c}{\textbf{Sampling probabilities}} 
& \multicolumn{6}{c}{\textbf{AudioCaps-Val}} 
& \multicolumn{4}{c}{\textbf{Commercial-Val}} \\

\cmidrule(lr){3-7} \cmidrule(lr){8-13} \cmidrule(lr){14-17}

&
& (*)
& \multicolumn{4}{c}{\mycaps} 
& \multicolumn{2}{c}{T2A} 
& \multicolumn{2}{c}{A2T-any}
& \multicolumn{2}{c}{A2T-all}
& \multicolumn{2}{c}{T2A} 
& \multicolumn{2}{c}{A2T}\\

\cmidrule(lr){4-7}
\cmidrule(lr){8-9} \cmidrule(lr){10-11}
\cmidrule(lr){12-13} \cmidrule(lr){14-15}
\cmidrule(lr){16-17}

&
& 
& M & R & RS & Tag
& R@5 & R@10 
& R@5 & R@10
& R@5 & R@10 
& R@5 & R@10
& R@5 & R@10 \\
\midrule

LAION \cite{wu2023large}
& AC+LA+AS+CL
& -- 
& -- & -- & -- & -- 
& 64.7 & 75.8
& 49.7 & 65.3
& 7.6 & 16.6 
& 23.5 & 34.8
& 28.7 & 41.6 \\

Ours$_{\textbf{AC+WC}}$
& AC+WC
& --
& -- & -- & -- & -- 
& 66.8 & 75.8
& 52.6 & 70.0
& 2.1 & 5.8
& 18.1 & 26.6
& 19.8 & 28.8 
 \\
\midrule

Ours$^{(1)}$
&(*)
& 1.0 
& -- & -- & -- & -- 
& 71.6 & 82.4
& 62.4 & 78.4
& 4.5 & 8.9 
& 23.6 & 32.6 
& 26.6 & 35.8 
 \\

Ours$^{(2)}$
& \mycaps
& -- 
& 1.0 & -- & -- & -- 
& 71.1 & 84.2
& 67.6 & 82.9
& 2.9 & 8.9 
& 23.0 & 32.7
& 24.4 & 33.4
\\

Ours$^{(3)}$
& (*)+\mycaps
& 0.8
& 0.2 & -- & -- & -- 
& 75.8 & 87.1
& 73.2 & 86.3
& 5.8 & 11.8
& 27.3 & 37.4
& 31.0 & 39.8 
\\

Ours$^{(4)}$
& (*)+\mycaps
& 0.2
& 0.8 & -- & -- & -- 
& 75.5 & 86.3
& 72.1 & 83.7
& 4.7 & 12.1 
& 27.3 & 36.0
& 28.0 & 38.5
\\

Ours$^{(5)}$
& (*)+\mycaps
& 0.2
& 0.4 & -- & 0.4 & -- 
& 76.6 & 87.4
& 68.7 & 85.0
& 8.7 & 17.1 
& 29.0 & 38.0
& 33.8 & 42.0
\\

Ours$^{(6)}$
& (*)+\mycaps
& 0.2
& 0.4 & -- & -- & 0.4 
& 75.5 & 87.1
& 71.6 & 87.4
& 6.8 & 16.6
& 29.3 & 39.1
& 33.7 & 42.8
\\

Ours$^{(7)}$
& (*)+\mycaps
& 0.2
& 0.4 & -- & 0.3 & 0.1 
& 77.6 & 89.5
& \textbf{72.9} & 86.3
& 9.5 & 15.5
& 31.1 & 41.1
& 34.6 & 44.3\\

Ours$^{(8)}$
& (*)+\mycaps
& 0.1
& 0.3 & 0.2 & 0.3 & 0.1
& 78.9 & 88.7
& 71.6 & \textbf{86.8}
& 8.4 & \textbf{19.7} 
& 31.1 & 41.4
& 35.3 & \textbf{45.0}
\\

Ours$^{(9)}$ = \textsc{\myclap$_{\text{AR}}$}
& (*)+\mycaps
& 0.2
& 0.2 & 0.2 & 0.2 & 0.2 
& \textbf{79.5} & \textbf{91.3}
& 71.3 & 86.3
& \textbf{9.5} & 19.5
& \textbf{32.4} & \textbf{43.2}
& \textbf{35.9} & 44.6
\\

\bottomrule
\end{tabular}
\par\vspace{1.5mm}
\begin{minipage}{0.80\linewidth}
\footnotesize
\end{minipage}
(*) and \mycaps \space share the same audio corpus, but have different captions (see \autoref{subsection:AR-retrieval}). “AC” refers to AudioCaps, “LA” refers to LAION-630K, “CL” refers to Clotho, and “AS” refers to AudioSet. 
\end{table*}

\subsection{Subjective Evaluation}

To assess the quality of audio captioning systems, we conducted a perceptual evaluation study. The evaluation corpus comprised 60 audio samples, partitioned into four subsets of 15 samples each, drawn from four different audio captioning datasets.
The audio and caption pairs were sampled from FreeSound (Raw and WavCaps captions), AudioSet (AudioCaps captions), AudioSet Strongly Labeled (WavCaps captions), and finally \mycaps \space (\texttt{main} captions). BBC Sound Effects was excluded due to missing shared audio identifiers across captioning datasets, preventing reliable caption alignment.

Each participant evaluated 15 samples randomly selected from this fixed pool of 60 audio excerpts. For each audio sample, participants were presented with two candidate captions ``Caption A'' and ``Caption B'', and asked to evaluate their quality according to the following criteria: 

\begin{itemize}
    \item \textit{Completeness}: Does the caption cover all sound events present in the audio?
    \item \textit{Correctness}: Do the described elements accurately correspond to the auditory content?
    \item \textit{Plausibility}: Are all described elements consistent with what can realistically be perceived in the audio? 
    \item \textit{Qualitativeness}: Does the caption include meaningful descriptive details (e.g. texture, rhythm, intensity, background) that help imagine the sound?
\end{itemize}

These criteria reflect the underlying assumption that a high-quality caption should faithfully represent the audio content with sufficient details, while avoiding both omissions of sound events and the introduction of inaccurate or implausible information. We prioritize subjective plausibility to better reflect the one-to-many nature of audio-text relationships, where multiple linguistic descriptions can faithfully represent the same audio content. This differs from other works that emphasize video- or context-grounded plausibility, requiring captions to match the specific recording context. 

Participants rated each caption using a 5-point Mean Opinion Score (MOS) scale ranging from 1 (\textit{Bad}) to 5 (\textit{Excellent}), based on the criteria above. In addition to numerical ratings, qualitative feedback was collected for each caption through multiple-choice checkboxes and optional free-text comments. To that end, we defined an \textit{a priori} taxonomy of qualitative observations covering common captioning issues we observed in 
existing datasets. These issues, presented in \autoref{tab:taxonomy_obs}, include missing details, unsupported content, insufficient descriptiveness, factual imprecisions, and captions unrelated to the audio. 

This additional qualitative assessment is essential for characterizing the relationship between the perceived caption quality and the observed factors. 
Finally, participants reported their overall confidence for each item using a 5-point scale (1 = very unsure, 5 = very confident). The complete evaluation session, including instructions and practice examples, required approximately 15 minutes per participant.

\subsection{Zero-shot Audio Classification}

To evaluate the informativeness and cross-modal alignment of our CLAP representations, we conduct a zero-shot audio classification experiment. Both audio signals and textual class labels are embedded into the shared audio-text representation space, and classification is performed by computing cosine similarities between the audio embedding and all candidate class embeddings. The predicted label corresponds to the class with the highest similarity score.

We compare three of our CLAP models against the LAION-CLAP baseline and our CLAP model trained on AudioCaps and WavCaps. Performance is evaluated using recall $R@k$ with $k \in \{1,5\}$. We first report results on the widely adopted ESC-50 \cite{piczak2015esc}, a dataset derived from FreeSound, containing 2k 5-seconds audio clips across 50 classes. However, both our model and all baselines are trained on data originating from FreeSound, which overlaps with ESC-50 and may lead to inflated performance estimates due to dataset leakage. To address this limitation, we further evaluate on FoleyBench \cite{dixit2026foleybench}, a fully disjoint dataset sourced from video content and not included in any training data used by our models or baselines. FoleyBench contains 5k 10-seconds clips spanning 81 categories and provides a broad and balanced coverage of sound categories defined by the Universal Category System (UCS)\footnote{https://universalcategorysystem.com/}, a taxonomy designed for professional sound design.

\section{Results}
\subsection{Audio Language Retrieval}

This section investigates the impact of caption diversity on cross-modal audio-text alignment. ~\autoref{tab:AR_results} reports text-to-audio and audio-to-text retrieval performance, together with a series of ablation studies investigating the impact of different caption mixtures used during CLAP training.

We first compare the LAION-CLAP and Ours$_{\text{AC+WC}}$ models, the latter being trained on AudioCaps and WavCaps using our CLAP architecture. 

Overall, results are mixed between the two models: Ours$_{\text{AC+WC}}$ achieves stronger T2A and A2T-any performance on AudioCaps-Val, while LAION-CLAP obtains better A2T-all results on AudioCaps-Val as well as stronger performance on Commercial-Val.
We hypothesize that this discrepancy stems from the LAION-630k dataset used in LAION-CLAP, which is substantially larger and more diverse. Its broader training distribution may improve generalization to the out-of-domain Commercial-Val set, whereas the more specialized training distribution of Ours$_{\text{AC+WC}}$ may instead provide an advantage on the in-domain AudioCaps-Val benchmark.

Then, we evaluate models trained on a single caption source, i.e, Ours$^{(1)}$, which is trained on the baseline dataset (*), and Ours$^{(2)}$, which is trained on the \texttt{main} captions from \mycaps. 
Both models surpass LAION-CLAP on AudioCaps-Val but overall fall short on Commercial-Val. Ours$^{(2)}$ occasionally outperforms Ours$^{(1)}$ on AudioCaps-Val but it results in slightly weaker performance on the commercial benchmark. This indicates that, \textit{despite the higher subjective quality of \mycaps \space captions (see \autoref{section:results_subjective_test}), caption quality alone results in only marginal retrieval improvements. It may also suggest a distribution shift in terms of audio and captions from AudioCaps-Val to the Commercial-Val dataset.}

We next investigate the effect of using multiple caption sources. Models Ours$^{(3)}$ and Ours$^{(4)}$ combine the baseline dataset (*) with the \texttt{main} captions at different sampling ratios ($(0.8, 0.2),(0.2, 0.8)$). Both models consistently outperform the previous baselines, showing that \textit{combining seemingly complementary caption sources is more effective than relying on either annotation style alone.}

We then investigate whether increasing linguistic diversity further improves retrieval. 
Ours$^{(5,6,7)}$ progressively increase the linguistic complexity of the supervision by introducing \texttt{rephrased-short} and \texttt{tags} captions before incorporating the more fine-grained  \texttt{rephrased}. Both caption type individually improves retrieval performance while combining them yields further gains, indicating that multiple description granularities provide richer supervision than any single caption style. Finally, Ours$^{(8)}$ and Ours$^{(9)}$ leverage the complete \mycaps \space collection, including the \texttt{rephrased} captions and achieve the best overall retrieval performance for equally weighted sources.

Overall, these experiments consistently show that caption diversity rather than caption quality alone, is a significant factor driving improvements in CLAP training. Since our models are trained on approximately the same amount of audio data as LAION-CLAP, the observed gains cannot be attributed to audio data scaling. Instead, \textit{we show that increasing the diversity and scale of captions substantially strengthens cross-modal alignment without requiring additional audio data.}

\begin{figure}[t]
    \centering
    \includegraphics[width=1\linewidth]{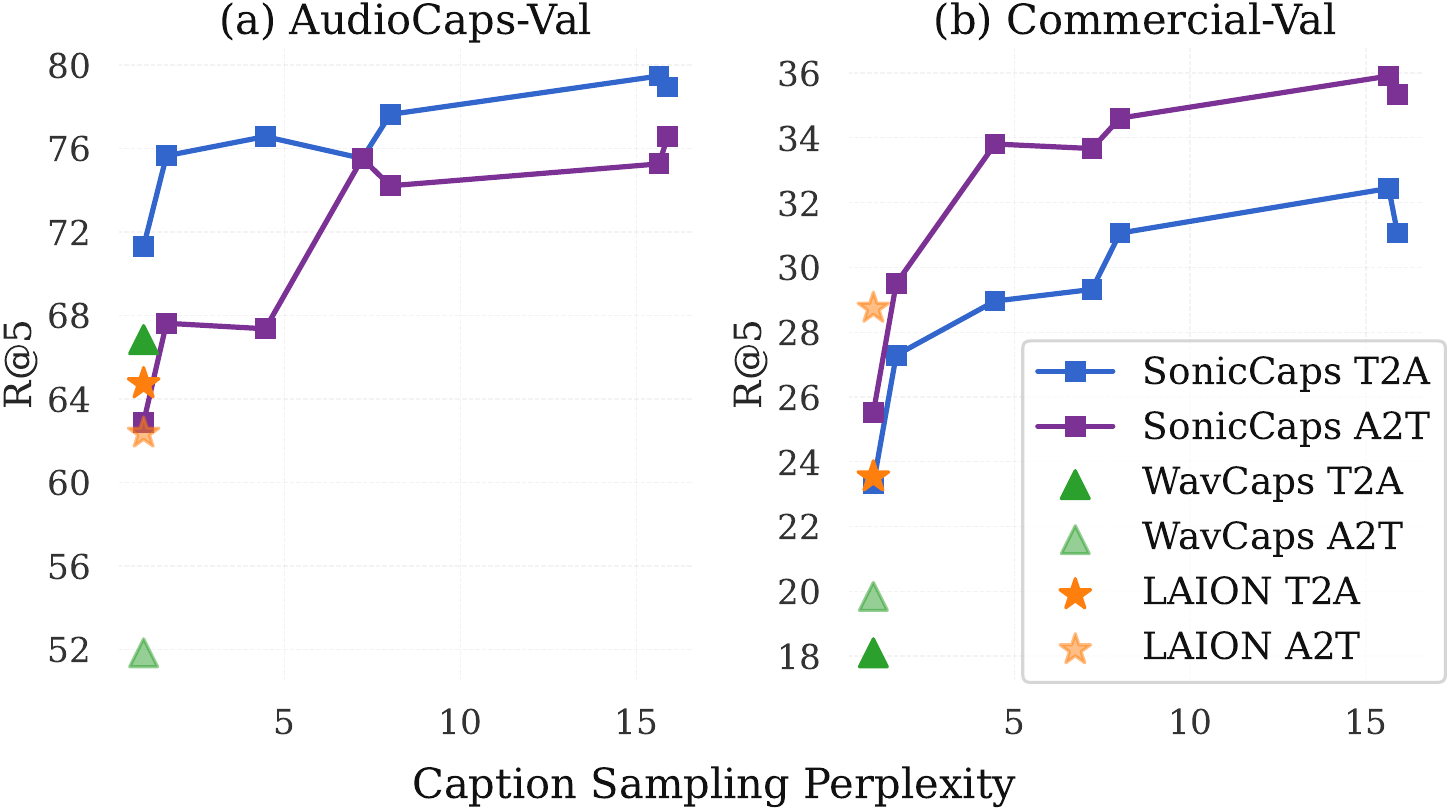}
    \caption{Text-to-Audio (T2A) and Audio-to-Text (A2T) R@5 Performance vs. Caption Sampling Perplexity (effective number of available captions per audio)  on (a) AudioCaps Validation Set and (b) Internal Validation Set.}
    \label{fig:ar-diversity}
\end{figure}

~\autoref{fig:ar-diversity} illustrates the effect of caption sampling perplexity, which can be interpreted as the effective number of captions available per audio sample during training. Under this definition, both WavCaps and LAION-CLAP are positioned at a perplexity of 1. Since several of our trained models, Ours$^{(i)}$, with $i \in \{1,\ldots,9\}$ share the same caption sampling perplexity, we report their average retrieval performance as a single point in the figure.
Note that, unlike ~\autoref{tab:AR_results}, A2T results on AudioCaps-Val are here computed in the standard setting, considering only a single caption among the five available per audio. The figure reveals a clear upward trend: performance improves as average caption diversity increases, consistently across both evaluation datasets and both A2T and T2A tasks. This provides a clear illustration of the benefit brought by the diversity of our dataset. Based on this evaluation, we release a specialized CLAP model for audio-language retrieval \myclap$_{\text{AR}}$ derived from Ours$^{(9)}$.

\subsection{Subjective Evaluation}
\label{section:results_subjective_test}

\begin{figure}[t]
    \centering
    \includegraphics[width=0.48\textwidth]{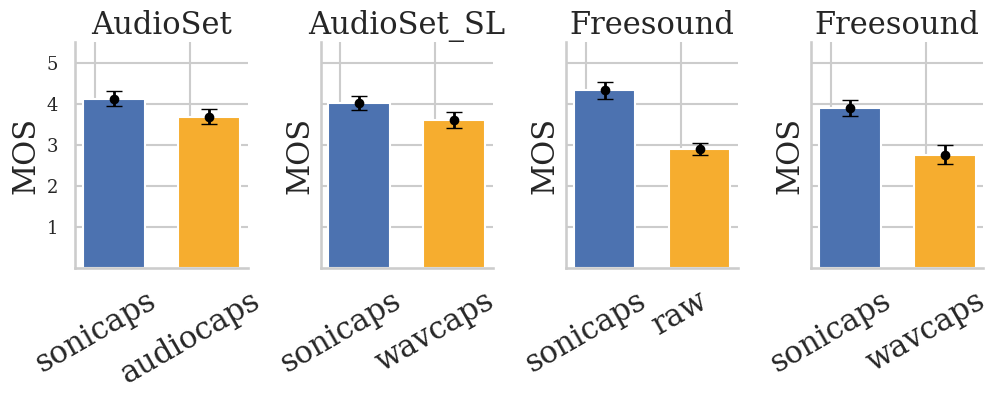}
    \caption{
    Subjective test results : Pairwise Mean opinion scores (MOS) across datasets, with 95\% confidence interval.}
    \label{fig:mos_results}
\end{figure}

A total of 25 participants took part in the subjective evaluation, yielding 375 assessments with an average of 6 ratings per item. Among them, $88\%$ reported familiarity with listening tests, and $50\%$ indicated prior experience with audio captioning. Evaluations with confidence scores below 3 were discarded, and items rated fewer than 3 times were excluded to ensure statistical reliability. We first report the mean opinion scores (MOS) derived from pairwise comparisons in \autoref{fig:mos_results}. Overall, \mycaps \space achieves the highest preference scores, outperforming human-authored captions from AudioCaps, raw FreeSound descriptions, as well as automatically generated captions from WavCaps. 

\begin{figure*}[t]
    \centering
    \includegraphics[width=0.9\textwidth]{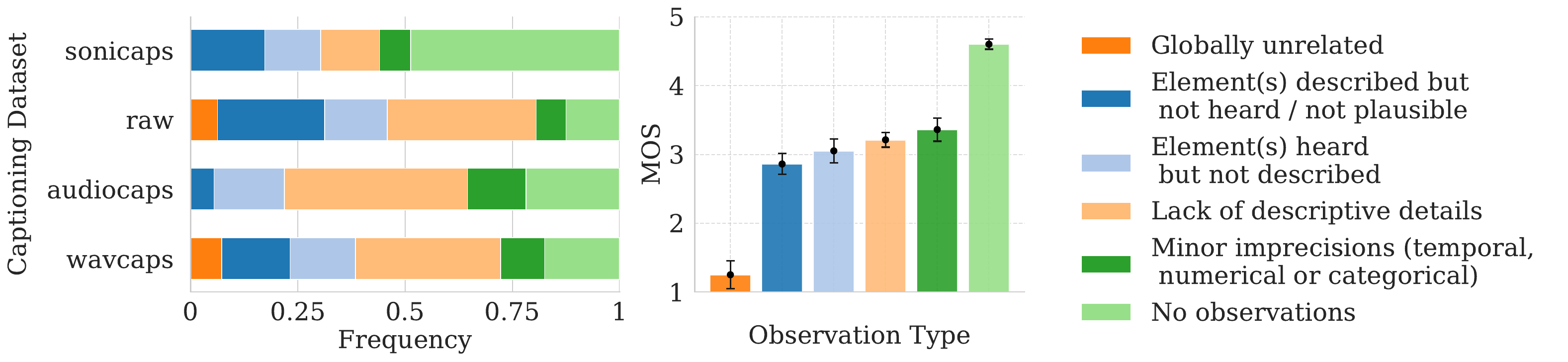}
    \caption{
    Subjective evaluation results: Distribution of annotated observations (left) across datasets and corresponding Mean Opinion Scores (right), averaged over all datasets.}
    \label{fig:obs_results}
\end{figure*}

\autoref{fig:obs_results} (left) shows the distribution of the five possible observation categories proposed during the evaluation, along with a sixth category ``No observations'' corresponding to evaluations where no observation box was checked. Several trends emerge from this analysis. Several trends emerge from this analysis. Globally unrelated captions are observed exclusively for WavCaps and Freesound Raw, indicating these are the only sources prone to substantial semantic misalignment between caption and audio content. Hallucinations and missing elements are comparable across datasets except for AudioCaps for which hallucinations are markedly less frequent, consistent with its human-annotated nature. Finally \mycaps \space stands out by producing substantially fewer captions lacking descriptive detail, with approximately 50\% receiving no negative observations at all, compared to less than 25\% for the other sources.

\autoref{fig:obs_results} (right) additionally report the average MOS as a function of the observation type. When no observation is made, the average score approaches the maximum rating, whereas hallucinations and missing elements appear to be the most penalizing factors. We note that these insights should be taken with caution, as the analysis does not account for the multi-factorial nature of the annotations, meaning that multiple observations can co-occur for a single caption.

Finally, we investigate to what extent CLAP-based similarity scores, as a proxy for text-audio alignment, correlate with human judgments. Since our subjective evaluation protocol is inherently pairwise, we consider two complementary notions of agreement between CLAP scores and MOS scores: (i) a pairwise agreement signal, based on score differences within each comparison, and (ii) an absolute agreement signal, based on absolute values assigned to each caption.
The pairwise formulation is the most reliable indicator in our setting, as it directly reflects the comparative nature of the evaluation task. In contrast, absolute MOS values are not  calibrated across annotators and should be interpreted with caution. We nonetheless report absolute correlations for completeness, as they provide an additional view of the alignment between model scores and human ratings. 

To this end, we compute the Spearman correlation $\rho(\Delta \text{MOS}, \Delta \text{CLAP})$, between MOS and CLAP score differences for each pair of evaluated captions, comparing LAION-CLAP and our model Ours$^{(2)}$. Caption pairs with zero MOS difference are excluded. We select Ours$^{(2)}$ for this analysis as it is trained on the \texttt{main} captions used in the subjective evaluation, enabling an in-domain comparison. \autoref{tab:spear-corr} shows that LAION-CLAP obtains a slightly negative correlation ($\rho=-0.07$), whereas Ours$^{(2)}$ achieves a substantially positive correlation ($\rho=0.32$), indicating significantly better alignment with human preferences. The same trend is observed when looking at the correlation between absolute MOS and CLAP scores $\rho(\text{MOS}, \text{CLAP})$, further confirming the robustness of the result. Based on this evaluation, we release a specialized CLAP model for perceptual caption quality assessment \myclap$_{\text{MOS}}$ derived from Ours$^{(2)}$.

\subsection{Zero-shot Audio classification}

\begin{table}[t]
    \centering
    \caption{Spearman correlation between MOS and CLAP score.}
    \begin{tabular}{l| c c}
    \Xhline{1pt}
    Model & $\rho(\Delta\text{MOS}, \Delta\text{CLAP})$ & $\rho(\text{MOS}, \text{CLAP})$\\
    \hline
    LAION-CLAP &  -0.07 & 0.03 \\
    Ours$^{(2)}$ = \myclap$_{\text{MOS}}$ &  \textbf{0.32} & \textbf{0.22}\\
    \Xhline{1pt}
    \end{tabular}
    \label{tab:spear-corr}
\end{table}

We report zero-shot classification results in \autoref{tab:zero_shot}, comparing a subset of our CLAP models with LAION-CLAP. Specifically, we evaluate Ours$_{\text{AC+WC}}$, Ours$^{(1)}$ and Ours$^{(2)}$, trained respectively on AudioCaps and WavCaps, the baseline dataset (*) and on \mycaps \space \texttt{main}, which was shown to better align with human judgments (\autoref{section:results_subjective_test}). We further include Ours$^{(8)}$ and Ours$^{(9)}$, selected for their superior audio-text retrieval performance.

\begin{table}[t]
\centering
\caption{Zero-shot classification results.}
\label{tab:zero_shot}
\small
\setlength{\tabcolsep}{3.8pt}

\begin{tabular}{lcc|cc}
\Xhline{1pt}

\multirow{2}{*}{Model} 
& \multicolumn{2}{c|}{ESC-50}  
& \multicolumn{2}{c}{FoleyBench} \\
\cline{2-5}

& R@1 & R@5 
& R@1 & R@5 
\\
\hline

LAION-CLAP & 82.1 & 93.6 & 1.34 & 9.14 \\
Ours$_{\textbf{AC+WC}}$ & 62.6 & 89.2 &  6.26 &  18.4\\
\hline
Ours$^{(1)}$ & 68.9 & 92.3 & 8.2 & 22.0 \\
Ours$^{(2)}$ & 65.6 & 89.9 & \textbf{9.66} & 24.8 \\
\hline
Ours$^{(8)}$ & \textbf{89.4} &  \textbf{99.2} & 8.80 & 22.4 \\
Ours$^{(9)}$ & 86.4 & 98.2 & 8.34 & \textbf{25.6} \\

\Xhline{1pt}
\end{tabular}
\end{table}

On ESC-50, LAION-CLAP outperforms our models trained with a single caption per audio (i.e., Ours$_{\text{AC+WC}}$ and Ours$^{(1,2)}$) However, this trend is reversed on FoleyBench, where all of our models consistently outperform LAION-CLAP, with gains of at least $10\%$ in $R@5$. Since this performance gap extends to Ours${_\text{AC+WC}}$, which is trained on a smaller and different audio corpus, the observed differences are not solely attributed to dataset scale and may also be explained by architectural choices. A more informative comparison is therefore obtained within our own models. Ours$^{(1)}$ outperforms Ours$^{(2)}$ on ESC-50 ($68.9\%$ vs. $65.6\%$ R@1), likely benefiting from the overlapping FreeSound data between the benchmark and the baseline dataset (*), in terms of both audio and captions.
The opposite behavior is observed on the fully disjoint FoleyBench, where Ours$^{(2)}$ surpasses Ours$^{(1)}$ ($9.66\%$ vs. $8.2\%$ R@1), suggesting that the \texttt{main} captions induce representations with stronger cross-domain transfer.
Increasing caption diversity reinforces this effect: Ours$^{(8)}$ and Ours$^{(9)}$, trained on the full \mycaps \space collection, obtain the best results on both benchmarks, with Ours$^{(9)}$ reaching $25.6\%$ R@5 on FoleyBench, compared to $18.4\%$ for Ours$_{\text{AC+WC}}$ and $9.1\%$ for LAION-CLAP.

Overall, these results provide further evidence that \textit{increasing caption diversity substantially improves the semantic representations learned by CLAP models, even when the underlying audio corpus remains unchanged}. The consistent gains on FoleyBench, whose audio content is entirely disjoint from the training data, suggest that \textit{richer linguistic supervision enhances performance and out-of-domain generalization}.

\section{Conclusion}

We introduce \mycaps, a new dataset of audio captions designed to improve both fidelity and semantic diversity. We study the impact of semantic caption diversity on learning audio–text representations within a CLAP framework. Through extensive ablations, we show that increasing caption diversity consistently improves cross-modal retrieval and zero-shot performance, outperforming both single-source and large-scale baselines, without having to scale audio data. In contrast, improvements in caption quality alone, validated by subjective evaluations, lead to only marginal gains in retrieval, highlighting diversity as the key driver of performance. However, higher-quality captions enable training CLAP models that are more aligned with human perceptual judgments, improving the correlation with subjective evaluations of caption quality. Overall, our results indicate that enhancing and diversifying textual supervision is a key factor for improving both retrieval performance and perceptual alignment in audio–text learning.

\section*{Acknowledgment}

We thank Pro Sound Effects for granting permission to release captions associated with their datasets, which made this work possible. We are also grateful to the Sony AI team members for their valuable guidance, discussions, and feedback throughout the project. In particular, we would like to acknowledge \textit{Alexandre Bittar},
\textit{Gaëtan Hadjeres}, \textit{Thomas Hummel}, \textit{Khaled Koutini}, 
\textit{Hakim Missoum}, \textit{Fabio Morreale}, \textit{Joan Serra} and \textit{Benno Weck} for their insightful input and support at various stages of this work.

\bibliographystyle{IEEEbib}
\bibliography{refs}
\end{document}